\documentclass[prd,superscriptaddress, showkeys,notitlepage, 11pt]{revtex4-1}
\usepackage{bm}
\usepackage{xcolor}
\usepackage[
left=3.00cm,
right=3.00cm,
top=2.50cm,
bottom=2.50cm
]{geometry}
\usepackage{amsmath}
\begin{document}

\title{FINITE AND DIVERGENT PARTS OF THE SELF-FORCE OF A POINT CHARGE FROM ITS SPHERICALLY AVERAGED SELF-FIELD} 
\author{V. Hnizdo}
\email{hnizdo2044@gmail.com}
\affiliation{2044 Georgian Lane, Morgantown, WV 26508, USA}
\author{G. Vaman}
\email{vaman@ifin.nipne.ro}
\affiliation{Institute of Atomic Physics, 30 Reactorului St., P.\,O.\,Box MG-6, Bucharest, Romania}
 
\begin{abstract}The electromagnetic self-force  of a point charge  moving arbitrarily on a rectilinear trajectory is calculated by averaging  its retarded electric self-field over a sphere of infinitesimal radius centered on the charge's present position. The finite part of the self-force obtained is the well-established relativistic radiation reaction, while its divergent part implies the pre-relativistic longitudinal electromagnetic mass of Abraham.  
\end{abstract}

\maketitle

\noindent {\bf 1. Introduction}

 Recently, we have obtained the relativistic Lorentz-Abraham-Dirac (LAD) radiation-reaction force \cite{dir} of a point charge moving arbitrarily on a rectilinear trajectory by calculating the negative of the time rate of change of the momentum of the charge's retarded self-field \cite{hni1}. Remarkably, the non-manifestly-covariant calculation procedure employed in \cite{hni1} involved no inertial-mass renormalization or any other explicit removal of infinities. The procedure thus amounted to the extraction of a finite part of the divergent integral for the point charge's retarded-self-field momentum.

In this paper, we revisit the same problem, but calculate the radiation-reaction force using the finite part of the  average of the charge's  retarded self-field over a sphere of infinitesimal radius centered on its present (current) position. 
The calculation, unlike that in \cite{hni1}, yields also the divergent part of the self-force. It will turn out that the electromagnetic mass implied by the divergent part is surprisingly the longitudinal electromagnetic mass that Abraham obtained in his theory of a non-Lorentz-contractible electron before the advent of special relativity, 
but the finite part is again the relativistic LAD radiation reaction for rectilinear motion. 

The procedure of calculating the radiation-reaction force of a point charge using  a suitable  average of its self-field has already been employed by several authors \cite{tei}--\cite{man}. Teitelboim's calculation \cite{tei} was manifestly covariant, yielding the full relativistic LAD equation of motion. Boyer \cite{boy} and Haque \cite{haq} used the procedure to obtain  the radiation-reaction force of a charge in nonrelativistic uniform circular motion.  Mansuripur \cite{man} employed the average of what is effectively half the difference of the retarded and advanced self-fields (following in this Dirac \cite{dir}) over a sphere of vanishingly small radius, centered on the charge's retarded position in its instantaneous rest frame.  The self-force thus obtained is the well-known nonrelativistic  Lorentz-Abraham radiation-reaction force, which on the  Lorentz transformation to the laboratory frame becomes the exact LAD radiation reaction. It should be noted that,  like the calculation in \cite{hni1},  Mansuripur's procedure did not require any renormalization of the charge's inertial mass.
\\

\noindent {\bf 2. Averaging the retarded electric self-field (Finite part)}

We consider a point charge $e$ moving arbitrarily on a rectilinear trajectory $w(t)$ along the $x$-axis.
It is convenient to assume that  $w(t)\hat{\bf x}$ vanishes at the given instant of time $t$, so that the charge is  located at the origin ${\bf r}=0$ of the coordinate system at that time instant. The radius vector $\bf r$ of observation will thus coincide with  the displacement from the position of the charge at the given present time $t$. 

The $x$-component of the charge's retarded electric field can be written as 
\begin{equation}
E_x({\bf r},t)= E_{1\,x}({\bf r},t) + E_{2\,x}({\bf r},t),
\end{equation}
where
\begin{align}
E_{1\,x}({\bf r},t) &=-\frac{\partial \phi({\bf r},t)}{\partial x} 
= -e \sum_{n=0}^{\infty} \frac{(-1)^n}{n!}\, \partial_x^{n+1} \frac{w^n(t-r/c)}{r}, \label{E1}\\
E_{2\,x}({\bf r},t)&=-\frac{\partial A({\bf r},t)}{c\,\partial t} 
= - \frac{e}{c^2} \frac{\partial}{\partial t} \sum_{n=0}^{\infty} \frac{(-1)^n}{n!}\,\partial_x^n \frac{w^n(t-r/c) \dot{w}(t-r/c)}{r} \label{E2} .\end{align}
Here, $\phi({\bf r},t)$ and $A({\bf r},t)\hat{\bf x}$ are the pertinent retarded scalar and vector potentials, respectively; 
$\partial_{x}^n$ denotes the partial differentiation with respect to $x$ of order $n$, $w^n(\cdot)\equiv [w(\cdot)]^n$ and the overdot indicates time differentiation.  The retarded  potentials used here were obtained in \cite{hni1} (see Eqs.\,(8) and (9) there) using the Taylor-series expansions of  delta-function expressions for the charge and current densities of a moving point charge.

To be able to calculate the average of the charge's retarded electric field over a sphere centered on the present position of the charge, we shall need the following formulas, holding for the time $t$ at which $w(t)\hat{\bf x}$ is assumed to vanish:  
\begin{align}
\frac{d^p}{dt^p}w^k(t) &= \left\lbrace \begin{array}{cc} 0,& p<k\\ k!\dot{w}^k(t),& p=k \end{array} \right. ,
\label{dpwk} \\ 
\frac{d^{k+1}}{dt^{k+1}}w^k(t)&=\frac{k(k+1)!}{2}\dot{w}^{k-1}(t) \ddot{w}(t), \label{dk1wk}\\
\frac{d^{k+2}}{dt^{k+2}} w^k(t) &= \frac{k(k-1)(k+2)!}{8} \dot{w}^{k-2}(t)\ddot{w}^2(t)+\frac{k(k+2)!}{6} \dot{w}^{k-1}(t)\dddot{w}(t). \label{dk2wk}
\end{align}
They can be inferred from the differentiation analogue of the multinomial theorem,
\begin{align}
\frac{d^m}{dt^m}f^k(t) &=\sum_{j_1+{\dots +}j_k=m} \frac{m!}{j_1!\cdot\cdot\cdot j_k!} f^{(j_1)}(t)\cdot\cdot\cdot f^{(j_k)}(t) \nonumber \\ 
&= \sum_{p=0}^{k} \binom{k}{k-p} f^p(t) \sum_{\substack{j_1+ \dots +j_{k-p}=m \\j_1,\dots ,\,j_{k-p}>0}} 
\frac{m!}{j_1! \cdot\cdot\cdot j_{k-p}!} f^{(j_1)}(t)\cdot\cdot\cdot f^{(j_{k-p})}(t), 
\label{par}
\end{align} written in the second line so that its application to  $f(t)=w(t)$ at the time when $w(t)\hat{\bf x}=0$ is facilitated.

Using the fact that  $w^n(t)\dot{w}(t)= (n{+}1)^{-1} dw^{n+1}(t)/dt$  and expanding  $w^n({t-}r/c)$ in Taylor series, the fields (\ref{E1}) and (\ref{E2}) can be written as  
\begin{align}
&E_{1\,x}({\bf r},t) =-e \sum_{n=0}^{\infty} \sum_{i=n}^{\infty} \frac{(-1)^{n+i}}{n!\; i!\; c^i} 
 (\partial_x^{n+1} r^{i-1})\frac{d^i}{dt^i} w^n(t),\label{e1}\\
& E_{2\,x}({\bf r},t)= -\frac{e}{c^2} \sum_{n=0}^{\infty} \sum_{i=n-1}^{\infty} 
\frac{(-1)^{n+i}}{(n+1)!\; i!\; c^i} (\partial_x^{n} r^{i-1})\frac{d^{i+2}}{dt^{i+2}} w^{n+1}(t),\label{e2}
\end{align}
where the lower limits of the sums over $i$ are adjusted in accordance with formula (\ref{dpwk}).
The derivatives $\partial_x^{n+1}r^{i-1}$ and $\partial_x^{n}r^{i-1}$ can be calculated using Faa di Bruno's formula \cite{faa}, 
\begin{align}
& \frac{d^m}{dx^m} g(f(x)) = \sum_{\substack{b_1+2b_2+ {\dots +} mb_m=m \\ b_1+{\dots +} b_m=k}}  \frac{m!}{b_1! \cdot\cdot\cdot b_m!} g^{(k)}(f(x))\left( \frac{f^{(1)}(x)}{1!} \right)^{b_1} 
\cdot \cdot \cdot \left( \frac{f^{(m)}(x)}{m!} \right)^{b_m}, 
\end{align}
by choosing  $f(x)=x^2{+}y^2{+}z^2$ ($y$ and $z$ thus being regarded as parameters) and $g(f)=f^{(i-1)/2}$; note that only the first two derivatives of thus defined $f(x)$ are different from zero: $f'(x)=2x$ and $f''(x)=2$. 
This way, we obtain 
\begin{align}
\frac{\partial^{n+1}}{\partial x^{n+1}} r^{i-1}&= 
\sum_{b=0}^{n/2}\frac{ (-1)^{b+1+n/2}2^{2b+1}(n+1)!} {(2b+1)! (n/2-b)!} \left(  \frac{1-i}{2}\right)_{b+1+n/2} r^{i-n-2b-3}  x^{2b+1},\: n\;{\rm even},\label{even}\\
\frac{\partial^{n+1}}{\partial x^{n+1}} r^{i-1}&= 
\sum_{b=0}^{(n+1)/2}\frac{ (-1)^{b+(n+1)/2}2^{2b}(n+1)!} {(2b)! ((n+1)/2-b)!} \left( \frac{1-i}{2}\right)_{b+(n+1)/2} r^{i-n-2b-2}  x^{2b},\: n\;{\rm odd}, \label{odd}
\end{align}
where $(\tfrac{1}{2}-\frac{i}{2})_{b+1+n/2}$,  etc.\, are the Pochhammer symbols (\cite{pru}, Appendix I).
The expressions for $\partial_x^{n} r^{i-1}$ are obtained by replacing $n$ in (\ref{even}) and (\ref{odd}) with $n-1$, and changing ``even" to ``odd" and {\it vice versa}. 
These expressions facilitate angular integration of  (\ref{e1}) and (\ref{e2}).  
Since 
\begin{align}
\int d\Omega\, x^{2b} = 4\pi \frac{r^{2b}}{2b+1}
\end{align} 
and the angular integral of $x^{2b+1}$ vanishes, we obtain
for the angular averages of $E_{1\,x}({\bf r},t)$ and $E_{2\,x}({\bf r},t)$:
\begin{align}
\bar{E}_{1\,x}( r,t) &= \frac{1}{4\pi}\int d \Omega\, E_{1\,x}({\bf r},t) \nonumber \\
&= -8 e \sum_{n=0}^{\infty} \sum_{i=2n+1}^{\infty} \frac{(-1)^{i+n} 
2^{2n}(n+1) \left( \frac{i}{2} \right)!\left( \frac{1-i}{2} \right)_{n+1}}{(2n+3)!\,i!\,\left( \frac{i}{2}-n-1 \right)!\,  c^i}\, r^{i-2n-3}\,\frac{d^i}{d t^i} w^{2n+1}(t), \label{OmegaE1} \\
\bar{E}_{2\,x}(r,t)&= \frac{1}{4\pi}\int d \Omega\, E_{2\,x}({\bf r},t) \nonumber \\
&= - e \sum_{n=0}^{\infty} \sum_{i=2n-1}^{\infty} \frac{(-1)^{i+n} 2^{2n}\left( \frac{i}{2}\right)!\left( \frac{1-i}{2} \right)_n}{(2n+1)(2n+1)!\, i!\left( \frac{i}{2}-n\right)!\, \,c^{i+2}}\, r^{i-2n-1}\,\frac{d^{i+2}}{d t^{i+2}} w^{2n+1}(t) . \label{OmegaE2}
\end{align}

In the limit $r \rightarrow 0$, only the terms with  
$1/r^2$, $1/r$ and $r^0$ can contribute to  angular averages (\ref{OmegaE1}) and (\ref{OmegaE2}). 
We consider first the finite parts of the limits $r \rightarrow 0$ of these averages, obtained accordingly by retaining only the terms with $i=2n+3$ and $i=2n+1$ in (\ref{OmegaE1}) and (\ref{OmegaE2}), respectively,  and using formulas (\ref{dk1wk}) and (\ref{dk2wk}):
\begin{align}
\bar{E}_{1\,x}^{\,\rm fin}(t) &={\rm f.p.}\lim_{r \rightarrow 0}\bar{E}_{1\,x}( r,t) \nonumber \\ 
&=  e \sum_{n=0}^{\infty}\frac{(-1)^n}{n!\,c^{2n+3}}(2n+1)(-n-1)_{n+1}\Big[\frac{n}{2}\dot{w}^{2n-1}(t)\ddot{w}^2(t)+\frac{1}{3}\dot{w}^{2n}(t)\dddot{w}(t)\Big],\\
\bar{E}_{2\, x}^{\,\rm fin}(t) &={\rm f.p.}\lim_{r \rightarrow 0}\bar{E}_{2\,x}(r,t) \nonumber \\
 &= e\sum_{n=0}^{\infty}\frac{(-1)^n}{n!\,c^{2n+3}}(n+1)(2n+3)(-n)_n\Big[\frac{n}{2}\dot{w}^{2n-1}(t) \ddot{w}^2(t)+ \frac{1}{3}\dot{w}^{2n}(t) \dddot{w}(t)\Big].
\end{align}
The series over $n$ can be summed using
\begin{align}
\sum_{n=0}^{\infty} \frac{(-1)^n}{n!} n(2n+1) (-n-1)_{n+1} \beta^{2n-1}&= -6\,\frac{ \beta (1+\beta^2)}{(1-
\beta^2)^4}, \\
\sum_{n=0}^{\infty} \frac{(-1)^n}{n!}(2n+1)(-n-1)_{n+1}\beta^{2n} &= -\frac{1+3\beta^2}{(1-\beta^2)^3}, \\
\sum_{n=0}^{\infty} \frac{(-1)^n}{n!}n(n+1)(2n+3)(-n)_n\beta^{2n-1}& =2\,\frac{\beta(5+\beta^2)}{(1-\beta^2)^4},
\\
\sum_{n=0}^{\infty} \frac{(-1)^n}{n!} (n+1)(2n+3)(-n)_n\beta^{2n} &=\frac{3+\beta^2}{(1-\beta^2)^3}, 
\end{align}
where $\beta=\dot{w}(t)/c$, and we obtain
\begin{align}
\bar{E}_{1\, x}^{\,\rm fin}(t) &=-3e \, \frac{\beta(1+\beta^2)}{(1-\beta^2)^4}\frac{\ddot{w}^2(t)}{c^4}- \frac{ e}{3}\,  \frac{1+ 3\beta^2}{(1-\beta^2)^3}\frac{\dddot{w}(t)}{c^3},\\
\bar{E}_{2\, x}^{\,\rm fin}(t) & = e \, \frac{\beta(5+\beta^2)}{(1-\beta^2)^4}\frac{\ddot{w}^2(t)}{c^4} + \frac{ e}{3}\, \frac{3+\beta^2}{(1-\beta^2)^3} \frac{\dddot{w}(t)}{c^3}.
\end{align}
The angular averages of the electric-field components $E_y$ and $E_z$, along with those of  the magnetic field, can be seen to vanish already on account of symmetry.  
The  retarded-self-field force of a point charge $e$ moving arbitrarily along the $x$-axis, calculated using the finite part of the self-field's average over a sphere of infinitesimal radius, centered on the charge's present position, is thus   
\begin{align}
{\bf F}_{\rm fin} &=e[\bar{E}_{1\,x}^{\,\rm fin}(t) +\bar{E}_{2\,x}^{\,\rm fin}(t)]\,\hat{\bf x} \nonumber \\
&= \frac{2e^2}{3c^3} \gamma^4\left[ \ddot{v}(t)+ \frac{3}{c^2} \gamma^2 v(t) \dot{v}^2(t)\right]\hat{\bf x},
\end{align}
where we now write the velocity $\dot{w}(t)$ as $v(t)$ and $\gamma=(1-v^2/c^2)^{-1/2}$. It is  exactly the same relativistic radiation-reaction force as that obtained in \cite{hni1}.
\\

\noindent{\bf 3. Averaging the retarded electric self-field (Divergent part)}

We now calculate the divergent parts of the fields (\ref{E1}) and (\ref{E2}) averaged over a sphere of infinitesimal radius centered on the charge. To do this,  we take $i=2n+1$, $2n+2$ and $2n-1$, $2n$  in Eqs.\ (\ref{OmegaE1}) and (\ref{OmegaE2}), respectively.  Since the Pochhammer symbols $(-n)_{n+1}$ and $(1-n)_{n}$ vanish, the resulting terms with $1/r^2$ vanish, too. Using formulas (\ref{dk1wk}) and (\ref{dk2wk}), we thus have for the divergent parts of the limits $r\rightarrow 0$ of (\ref{OmegaE1}) and (\ref{OmegaE2}):
\begin{align} 
 \bar{E}_{1\,x}^{\,\rm div}(t)&= -\lim_{r \rightarrow 0} \left( \frac{4 e\ddot{w}(t)}{rc^2} \right) 
\sum_{n=0}^{\infty} \frac{(-1)^n 2^{2n}(n+1)(2n+1)(n+1)! \left( -n-\frac{1}{2} \right)_{n+1}}{(2n+3)!\,c^{2n}}\,\dot{w}^{2n}(t),\\
 \bar{E}_{2\,x}^{\,\rm div}(t)&= -\lim_{r \rightarrow 0} \left( \frac{ e \ddot{w}(t)}{rc^2} \right)
\sum_{n=0}^{\infty} \frac{(-1)^n 2^{2n}(n+1)!\left(\frac{1}{2}-n \right)_n}{(2n)!\, c^{2n}}\,\dot{w}^{2n}(t).
\end{align}
Using now
\begin{align}
\left(-n-\frac{1}{2}\right)_{n+1}=(-1)^{n+1}\frac{(2n+1)!}{2^{2n+1}n!},\;\;\;\; 
\left( \frac{1}{2}-n \right)_n=(-1)^n\frac{(2n)!}{2^{2n}n!},
\end{align}
and summing the series over $n$,
\begin{align}
\sum_{n=0}^{\infty} \frac{(n+1)(2n+1)}{2n+3}\beta^{2n}&=
\gamma^4 +\frac{1}{\beta^3}\left(\frac{1}{2}\ln\frac{1+\beta}{1-\beta}-\frac{\beta}{1-\beta^2}\right),\\
\sum_{n=0}^{\infty} (n+1) \beta^{2n}&=\gamma^4,
\end{align}
where $\gamma=(1-\beta^2)^{-1/2}$ and $\beta =\dot{w}(t)/c$,
we obtain for the divergent part of the force of the electric self-field averaged over a sphere of infinitesimal radius: 
\begin{align}
{\bf F}_{\rm div}&=e[\bar{E}_{1\,x}^{\,\rm div}(t)+\bar{E}_{2\,x}^{\,\rm div}(t)]\,\hat{\bf x}
\nonumber \\
\label{Fdiv}
&= \lim_{r\rightarrow 0}\left(\frac{e^2 \dot{\beta}}{rc}\right)
\frac{1}{\beta^3}\left(\frac{1}{2}\ln\frac{1+\beta}{1-\beta}-\frac{\beta}{1-\beta^2}\right)\hat{\bf x}.
\end{align}

Before the advent of special relativity, the so-called longitudinal electromagnetic mass, $m_{\rm{em}\,||}$, was taken to be the proportionality coefficient of the negative of the electromagnetic self-force proportional to the acceleration of a charge in rectilinear motion (\cite{lor}, sec.\ 27). At a distance $r=a$ from the charge's location, this coefficient is according to Eq.\ (\ref{Fdiv}): 
\begin{align}
\label{m}
m_{\rm{em}\,||}=\frac{e^2}{ac^2}
\frac{1}{\beta^3}\left(\frac{\beta}{1-\beta^2}-\frac{1}{2}\ln\frac{1+\beta}{1-\beta}\right).
\end{align} 
Perhaps surprisingly, it happens to equal the mass $m_{\rm{em}\,||}$ that Abraham obtained in his model of the electron as a uniformly charged ``rigid" (meaning today non-Lorentz-contractible) spherical shell of radius $a$ (\cite{abr}, p.\! 191, Eq.\ (117); see also \cite{lor}, p.\! 39, Eq.\ (68)).  

We note here that Abraham's electromagnetic mass $m_{{\rm em}\,||}$ and 
the longitudinal electromagnetic mass $2e^2\gamma^3/(3ac^2)$, obtained using the time derivative of the momentum $(2e^2\gamma/3ac^2){\bf v}$ of the well-known Heaviside fields of a uniformly moving charged shell, in its rest frame spherical with radius $a$  \cite{hni2}, agree only when $\beta<<1$.  
But it is interesting to note also that 
it can be shown (see Appendix) that when the divergent self-field momentum $\bf G$ of a uniformly moving point charge is evaluated by approaching the field-momentum density's singularity on a spherical surface of radius 
$a$, one obtains a rather more complicated expression:
\begin{align}
{\bf G} =\lim_{a\rightarrow 0}\frac{e^2}{8a c^2}\frac{1}{\beta^3}[(1+2\beta^2)\beta-(1-4\beta^2)\gamma\, 
  \rm{arctg}(\beta\gamma)]{\bf v}.
\end{align}
The ``electromagnetic-mass" factor multiplying here the velocity $\bf v$ reduces to the value 
$2 e^2\gamma/(3ac^2)$ only when $\beta<<1$. 
\\   

\noindent {\bf 4. Concluding remarks}

In a recent Comment \cite{hni3} on the derivation of the LAD equation of motion by Dondera \cite{don}, we have advanced a conjecture concerning the Hadamard decomposition of the retarded-self-field momentum of an arbitrarily moving point charge. According to the conjecture, the divergent electromagnetic mass implied by the self-field momentum of a uniformly moving point charge is given by 
$\lim_{\varepsilon \rightarrow 0}2 e^2\gamma/(3\varepsilon c^2)$.
The requisite integration of the divergent integral for the retarded-self-field momentum of such a charge is thus  understood there to be done so that the singularity in the self-field-momentum density is approached  on an oblate-spheroidal surface whose shape is congruent with that of a Lorentz-contracted sphere. 

Curiously, Rohrlich has attributed to Abraham an equation of motion (\cite{ror}, Eqs.\! (2.7)--(2.8)) that implies the same decomposition of the retarded-self-field momentum as our conjecture would for a charged 
rest-frame-spherical shell of radius $a$. Rohrlich's attribution is correct for the equation's  radiation-reaction force, which Abraham was indeed the first to obtain in its full relativistic form (\cite{abr}, p.\! 123, Eq.\ (85)), but not for its electromagnetic-inertia term since Abraham's ``rigid"  shell of charge does not undergo the Lorentz contraction when it is moving.
Yaghjian has also attributed the same equation of motion (\!{\it sans} the ``bare-mass" term) to  Abraham, but he has attributed it to Lorentz as well, saying that it was for a ``relativistically rigid" (i.e.\! Lorentz-contractible, in Lorentz's parlance ``deformable") spherical shell of charge (\cite{yag}, pp.\! 11-12). 

Using a spherical average of the retarded electric self-field of a point charge moving arbitrarily on a rectilinear trajectory, we obtained in this paper both the finite and divergent parts of its electromagnetic self-force. The finite part obtained is the same as the well-established relativistic LAD radiation reaction, but the longitudinal electromagnetic mass implied by the divergent part differs at relativistic velocities  markedly from that obtained from the electromagnetic momentum of a uniformly moving rest-frame-spherical shell of charge. This fact indicates that, in contrast to the divergent part of the self-force, the force of radiation reaction is not sensitive  to the details of  the models for the charge used and the methods of calculation employed.  

Abraham obtained his longitudinal electromagnetic mass as the time derivative of the electromagnetic momentum he calculated  for a uniformly moving ``rigid" spherical shell of charge, the exterior electric field of which is not the 
well-known Heaviside electric field of a uniformly moving  spheroidal, Lorentz-contracted shell of charge. The fact that  our divergent part of the self-force implies Abraham's electromagnetic mass is indeed surprising.

In closing, we should  caution that electromagnetic mass is today a concept of limited applicability and usefulness. It is rooted in the electromagnetic program of the beginning of the 20th century, in which Abraham and some other prominent physicists attempted to supplant classical mechanics by a field theory based on Maxwell's electromagnetism and the ether as the medium sustaining it. But after the advent of special relativity the program became rather quickly outmoded, and its key concepts, the ether  and  electromagnetic mass, superseded (an illuminating  account of this interesting episode in the history of physics is given in \cite{jan}).   
\\

\noindent{\bf Appendix}

The well-known  Heaviside expressions for the fields of a point charge $e$ moving with a constant velocity 
$\bf v$ are given by
\begin{align}
{\bf E}({\bf r},t) = \frac{e}{\gamma^2(1-\beta^2 \sin^2\theta)^{3/2}} \frac{\bf r}{r^3},\;\;\;\;
{\bf B}({\bf r},t)= ({\bm \beta} \times {\bf E}),
\end{align}
where ${\bm \beta} ={\bf v}/c$, $\gamma=(1-\beta^2)^{-1/2}$ and $\theta$ is the angle between $\bf v$ and $\bf r$, the latter being the displacement from the present position of the charge. The momentum of these fields  is 
\begin{align} 
{\bf G}&= \frac{1}{4\pi c}\int d^3 r\, ({\bf E} \times {\bf B}) \nonumber \\
&=\frac{1}{4\pi c}\int d^3 r\, (1-\cos^2\theta)E^2{\bm\beta},
\end{align}
where account is taken of the fact that only the component  
of the vector product $({\bf E} \times {\bf B})$ along the direction of $\bm \beta$ contributes to the integral.
The square of the Heaviside electric field can be written as 
\begin{align}
E^2 = \frac{\gamma^2}{(1 +\beta^2\gamma^2\cos^2\theta)^3}\,\frac{e^2}{r^4},
\end{align}
and thus the divergent momentum $\bf G$ calculated by approaching the singularity at $r=0$ on a spherical surface of radius $a$ is given by
\begin{align}
{\bf G} &=\lim_{a\rightarrow 0}\frac{e^2\gamma^2}{4\pi c}\int_a^{\infty}\frac{dr}{r^2}\,
\int  \frac{d\Omega\,(1-\cos^2\theta)}{(1 +\beta^2\gamma^2\cos^2\theta)^3}\,{\bm \beta}\nonumber \\
&= \lim_{a\rightarrow 0}\frac{e^2\gamma^2}{2 a c }\int_{-1}^1 d\xi\, 
\frac{(1-\xi^2)}{(1+\beta^2\gamma^2\xi^2)^3}\, {\bm \beta}\nonumber \nonumber \\
&=\lim_{a\rightarrow 0}\frac{e^2}{8a c^2}\frac{1}{\beta^3}[(1+2\beta^2)\beta-(1-4\beta^2)\gamma\, 
  \rm{arctg}(\beta\gamma)]\,{\bf v}.
\end{align}

\end{document}